\documentstyle[multicol,psfig,prl,aps]{revtex}
\begin{document}
\draft

\title{Quantum chiral phases in frustrated easy-plane spin chains}

\author{A. K. Kolezhuk\cite{email}}
\address{Institut f\"ur Theoretische Physik, Universit\"at Hannover,
Appelstra{\ss}e 2, D-30167 Hannover, Germany\\
and Institute of Magnetism, National Academy of  
Sciences and Ministry of Education of Ukraine\\
 36(b) Vernadskii avenue, Kiev 03142, Ukraine}
\date{\today}

\maketitle

\begin{abstract}
The phase diagram of antiferromagnetic spin-$S$ chain with XY-type
anisotropy and frustrating next-nearest-neighbor interaction is
studied in the limit of large integer $S$ with the help of a
field-theoretical approach. It is shown that the existence of gapless
and gapped chiral phases found in recent numerical studies [M.Kaburagi
et al., J. Phys. Soc. Jpn. {\bf 68}, 3185 (1999), T.Hikihara et al.,
J. Phys. Soc. Jpn. {\bf 69}, 259 (2000)] is not specific for $S=1$,
but is rather a generic large-$S$ feature.  Estimates for the
corresponding transition boundaries are obtained, and a sketch of the
typical phase diagram is presented.  It is also shown that frustration
stabilizes the Haldane phase against the variation of the anisotropy.
\end{abstract}


\begin{multicols}{2}

In recent few years, the problem of possible nontrivial ordering in
frustrated quantum spin chains has attracted considerable attention
\cite{NersGogEss98,AligBatEss00,AllenSen99,Kab+99,Hik+00}.  Nersesyan
{\em et al.} predicted \cite{NersGogEss98} that in anisotropic
(easy-plane) antiferromagnetic $S={1\over2}$ chain with sufficiently
strong frustrating next-nearest-neighbor (NNN) coupling, a new phase
with a broken parity appears, which is characterized by the nonzero
value of {\em chirality}
\begin{equation}
\label{chirality} 
\kappa_{n}^{z} \equiv \langle({\mathbf S}_{n}\times{\mathbf
S}_{n+1})_{z} \rangle\,;
\end{equation}
note that the definition (\ref{chirality}) differs from the other,
so-called scalar chirality $\widetilde{\kappa}\propto {\mathbf
S}_{n-1}\cdot ({\mathbf S}_{n}\times{\mathbf S}_{n+1})$ which is often
discussed in the context of the isotropic spin chains
\cite{Zvyagin+Frahm}.  This prediction was made on the basis of the
bosonization technique combined with a subsequent mean-field-type
decoupling procedure. A similar conclusion can be reached
\cite{Bouzerar-Brenig} by means of a mean-field decoupling of the
quartic terms in the Jordan-Wigner transformed fermionic version of
the Hamiltonian, in the spirit of the Haldane's treatment of
spontaneously dimerized phase \cite{Haldane82}. Up to now, however,
this prediction for $S={1\over2}$ was not confirmed in numeric studies
\cite{Kab+99,AligBatEss00}. On the other hand, {\em two\/} different
types of chiral ordered phases, gapped and gapless, were found
numerically in $S=1$ easy-plane frustrated chain \cite{Kab+99,Hik+00}.
At present, to our knowledge, there is no theoretical analysis
addressing the problem of chiral ordered phases for the $S\geq 1$
case.

The aim of the present Letter is to study the generic 
large-$S$ behavior of antiferromagnetic easy-plane integer-$S$ chain
with frustrating NNN interaction, described by the following
Hamiltonian:
\begin{equation} 
\label{ham} 
\widehat{H}=J\sum_{n}\{ ({\mathbf S}_{n}{\mathbf S}_{n+1})_{\Delta} +
j({\mathbf S}_{n}{\mathbf S}_{n+2})_{\Delta} + D(S_{n}^{z})^{2} \}\,,
\end{equation}
Here $({\mathbf S}_{1}{\mathbf S}_{2})_{\Delta}\equiv
S_{1}^{x}S_{2}^{x} + S_{1}^{y}S_{2}^{y} +\Delta S_{1}^{z}S_{2}^{z}$,
${\mathbf S}_{n}$ denotes the spin-$S$ operator at the $n$-th site,
the lattice spacing $a$ has been set to unity, $J>0$ is the
nearest-neighbor exchange constant, $j>0$ is the relative strength of
the NNN coupling, and $0<\Delta<1$ and $D>0$ are respectively the dipolar
(inter-ion) and the single-ion anisotropies.

We argue that the existence of gapless and gapped chiral phases
found in \cite{Kab+99,Hik+00} is not specific for $S=1$, but is rather
a generic large-$S$ feature. Estimates for the corresponding
transition boundaries are obtained, and a sketch of the typical phase
diagram is presented.  As a side result, we also show that the domain
of stability of the Haldane phase against the anisotropy variation 
grows when the frustrating coupling $j$ is
increased, in accordance with the numerical results
\cite{Tonegawa+95}.

We use the well-known technique of spin coherent states
\cite{reviews1,reviews2} 
which effectively replaces spin operators by classical vectors
$(S_{n}^{+},S_{n}^{z})=S(\sin\theta_{n}e^{i\varphi_{n}},\cos\theta_{n})$
and incorporates the quantum features by means of
the path integral over all space-time
configurations of $(\theta,\varphi)$. The classical ground
state of (\ref{ham}) is well-known: the spins always lie in the easy
plane $(xy)$, i.e. $\theta={\pi\over2}$, for $j<{1\over4}$ the
alignment of spins is antiferromagnetic, $\varphi_{n}=\varphi_{0} +\pi
n$, and for $j>{1\over4}$ a helical structure with incommensurate
magnetic order develops, $\varphi_{n}=\varphi_{0}\pm (\pi-\lambda_{0})
n$, where $\lambda_{0}=\arccos(1/4j)$, and the $\pm$ signs above
correspond to the two possible chiralities of the helix.

In one dimension the long-range helical ordering is impossible since
it would imply a spontaneous breaking of the continuous in-plane
symmetry; in contrast to that, the existence of the finite chirality
$\kappa_{n}^{z}=\langle \sin(\varphi_{n+1}-\varphi_{n})\rangle$ is
not prohibited by the Mermin-Wagner theorem.

The classical {\em isotropic\/} system has for $j>{1\over4}$ three
massless modes with wave vectors $q=0$, $q=\pm \delta$, where
$\delta\equiv\pi-\lambda_{0}$ is the pitch of the helix. The effective
field theory for the isotropic case is the so-called $SO(3)$ nonlinear
sigma model, with the order parameter described by the local rotation
matrix \cite{RaoSen94,AllenSenechal95}. The physics becomes simpler in
presence of anisotropy since the modes with $q=\pm \delta$
acquire a finite mass.  Our starting point will be the following
ansatz for the angular variables $\theta,\varphi$:
\begin{eqnarray} 
\label{ansatz} 
\theta_{n}&=&\pi/2+p_{n}+(\xi_{n}e^{i\delta
n}+\xi_{n}^{*}e^{-i\delta n})/2\nonumber\\
\varphi_{n}&=&\pi n +\psi_{n}+(w_{n}e^{i\delta
n}+w_{n}^{*}e^{-i\delta n})/2\,.
\end{eqnarray}
We assume that the fluctuations $p$, $\xi$, $w$ are small and that
they are smooth functions of $n$, slowly varying over the
characteristic distance $l_{0}=2\pi/\lambda_{0}$; the same property is
assumed for the function $\lambda_{n}\equiv\psi_{n+1}-\psi_{n}$ which
can be viewed as a dual variable to $\psi$ \cite{Harada84}.  

 After passing to the continuum in the effective Lagrangian $ L=\int
dx {\cal L}=\hbar S\sum_{n}(1-\cos\theta_{n})\partial_{t}\varphi_{n} -
H(\theta,\varphi)$ we average over $l_{0}$, making the oscillating
terms disappear. The resulting expression for ${\cal L}$ is
\begin{eqnarray} 
\label{L} 
{\cal L}&=&\hbar S \big\{ p(\partial_{t}\psi)(1-|\xi|^{2}/4) +
[w^{*}(\partial_{t}\xi) + w(\partial_{t}\xi^{*})]/4\big\} \nonumber\\
&-&JS^{2}\big\{ V[\lambda](1-|\xi|^{2}/2) + A_{0}p^{2}
+(A_{1}/2)|w|^{2}\big\} \nonumber\\
&-&(JS^{2}/4)\big\{ M |\xi|^{2} + F\Delta |\partial_{x}\xi|^{2}\big\}\,,
\end{eqnarray}
where the following notation has been used:
\begin{eqnarray} 
\label{notation} 
&& V[\lambda]=j\cos2\lambda-\cos\lambda -U_{0}
+(j/2)\cos2\lambda(\partial_{x}\lambda)^{2}, \nonumber\\
&& U_{0}=j\cos2\lambda_{0}-\cos\lambda_{0},\quad
F=\cos\lambda_{0}-4j\cos2\lambda_{0}, \nonumber\\
&& A_{0}=D+\Delta(1+j)-U_{0},\quad M=2[D-(1-\Delta)U_{0}], \nonumber\\ 
&& A_{1}=\cos^{2}\lambda_{0}+j\cos^{2}2\lambda_{0} -U_{0}\,.
\end{eqnarray}
Integrating out the ``slave'' fields $p\simeq
-(\hbar/2JSA_{0})\partial_{t}\psi$, $w\simeq
(\hbar/2JSA_{1})\partial_{t}\xi$, and passing to the imaginary time
$y=ict$, $c=JS(2A_{0})^{1/2}/\hbar$, we obtain the effective Euclidean
action
\begin{eqnarray} 
\label{AE1} 
{\cal A}_{E}&=&{1\over g_{0}}\int\int dx\, dy \Big\{
{1\over2}(\partial_{y}\psi)^{2} +V[\lambda]\Big\}
\Big(1-{1\over2}|\xi|^{2}\Big) \nonumber\\
&+&{\widetilde{E}\over 4g_{0}}\int d^{2}X \big\{
(\partial_{\mu}\xi^{*})(\partial_{\mu}\xi)+m_{0}^{2}|\xi|^{2} \big\},
\end{eqnarray}
where $(X_{1},X_{2})=(x,ic't)$, $c'=JS(A_{1}F\Delta)^{1/2}/\hbar$,
$\partial_{\mu}=\partial/\partial X_{\mu}$, and the constants $g_{0}$,
$\widetilde{E}$, $m_{0}$ are given by
\begin{equation} 
\label{bare}
g_{0}={\sqrt{2A_{0}}\over S},\quad 
\widetilde{E} =\left({A_{0}F\Delta \over A_{1}}\right)^{1/2},\quad
m_{0}^{2}={M\over F\Delta}\,.
\end{equation}
Further, integrating out the massive $\xi$ yields the effective action
for $\psi$ only, with a renormalized coupling $g_{\rm eff}$:
\begin{eqnarray} 
\label{AE} 
{\cal A}_{E}&=&{1\over g_{\rm eff}}\int\int dx\, dy \Big\{
{1\over2}(\partial_{y}\psi)^{2} +V[\lambda]\Big\}\,, \nonumber\\
g_{\rm eff}&=&
g_{0}/\big(1-{g_{0}\over 2\pi
\widetilde{E}}\ln(1+\Lambda^{2}/m_{0}^{2})\big) \,,
\end{eqnarray}
where $\Lambda=\pi$ is the lattice cutoff.
A similar derivation may be carried out for $j<1/4$: starting
from the ansatz of the type (\ref{ansatz}) with {\em real\/} $\xi$,
$w$ and $\delta=\pi$,  one arrives at the same result
(\ref{AE}), but with $\lambda_{0}$ set to zero in all quantities
defined in (\ref{notation}), (\ref{bare}).

We have mapped the initial quantum 1D model to the 2D classical XY
helimagnet at effective ``temperature'' $g_{\rm eff}$ described by the
effective action (\ref{AE}). The validity of this mapping is
determined by the requirements $g_{0}\ll 1$, $|w| \ll |\xi| \ll 1$,
$p\ll1$, which translate into
\begin{equation} 
\label{validity} 
S\gg (2A_{0})^{1/2},\quad \Lambda e^{-\pi\widetilde{E}/g_{0}}\ll m_{0}\ll
Sg_{0}/\widetilde{E}
\end{equation}
The first inequality above means that we are not allowed to consider
large $j\gtrsim S^{2}/2$, and the second one requires the anisotropy to
be within a certain range. We will be mainly interested in the
behavior of (\ref{AE}) for $j$ close to the Lifshitz point
$j_{L}\equiv{1\over4}$, then the condition for the anisotropy
transforms into 
\begin{equation} 
\label{anis} 
\zeta \pi^{2} \varepsilon e^{-2\pi S\sqrt{\zeta\varepsilon}}\ll 
3\mu/8\ll 1\,,
\end{equation}
where 
$\mu\equiv 1-\Delta+4D/3$, $\varepsilon\equiv |j-j_{L}|$, 
and the constant $\zeta=1$ for $j<j_{L}$ and $\zeta=2$ for $j>j_{L}$,
respectively. 

The model (\ref{AE}) possesses two basic types of
topological defects \cite{GarelDoniach80}: (i) domain walls connecting
regions of opposite chirality, and (ii) vortices, existing inside the
domains with certain chirality and destroying the long-range helical
magnetic order (only quasi-long-range helical order is possible at
finite $g_{\rm eff}$). Thus one may expect two phase transitions:
Ising-type transition (``freezing'' of the domain walls) which
corresponds to the onset of the chiral order, and the
Kosterlitz-Thouless (KT) transition (vortex unbinding) corresponding
to the transition from the gapless chiral phase with algebraically
decaying helical magnetic correlations $\langle
\cos\psi(x)\cos\psi(0)\rangle\propto (1/x)^{g_{\rm eff}/2\pi}$ to the
gapped chiral phase with only short-range helical order (but still
with the long-range chiral order $\langle
\kappa^{z}(x)\kappa^{z}(0)\rangle \to \text{const}$, $x\to\infty$).
For the two transitions to be possible, one has to assume (later this
assumption will be checked self-consistently) that the critical
temperature of the Ising-type transition is {\em higher\/} that the
corresponding temperature of the KT transition.

Inside the phase with broken chiral symmetry one can set
$\psi=\pm\lambda_{0}x+\phi_{\pm}$, then $\lambda\simeq
\pm\lambda_{0}+\partial_{x}\phi_{\pm}$ and $V[\lambda]\simeq
{1\over2}F(\partial_{x}\phi_{\pm})^{2}$. One obtains then the
following estimate for the KT temperature:
\begin{equation} 
\label{gKT} 
g_{c}^{KT}\simeq (\pi/2)\sqrt{F}\,.
\end{equation}
The equation $g_{\rm eff}=g_{c}^{KT}$ determines the transition from
the chiral gapless to the chiral gapped phase at $j>j_{L}$, as well as
the transition from the non-chiral ($\lambda_{0}=0$) gapless XY phase
to the non-chiral gapped Haldane phase at $j<j_{L}$.  Note that
(\ref{gKT}) is still valid away of the Lifshitz point $j=j_{L}$, since
the field $\phi$ remains smooth far from the vortex core, and the KT
transition temperature is determined by the logarithmic divergence in
the free vortex energy at large distances.

In order to estimate the critical temperature of the Ising transition,
let us first make some observations concerning the properties of
chiral domain walls. The domain wall (DW) energy can be easily
calculated in the vicinity of the Lifshitz point, where
$\lambda\ll 1$, so that the potential $V[\lambda]\simeq (1/8)\{
(\lambda^{2}-\lambda_{0}^{2})^{2}+(\partial_{x}\lambda)^{2}\}$ takes
the form of the $\varphi^{4}$ model, and one readily obtains the static DW
solution $\lambda=\lambda_{0}\tanh
\big\{\lambda_{0}(x-x_{DW})\big\}$ and the corresponding
energy (per unit length in the $y$ direction)
\begin{equation} 
\label{Edw} 
E_{DW}\simeq \lambda_{0}^{3}/3,\qquad j-j_{L} \ll 1\,.
\end{equation}

\begin{figure}
\mbox{\psfig{figure=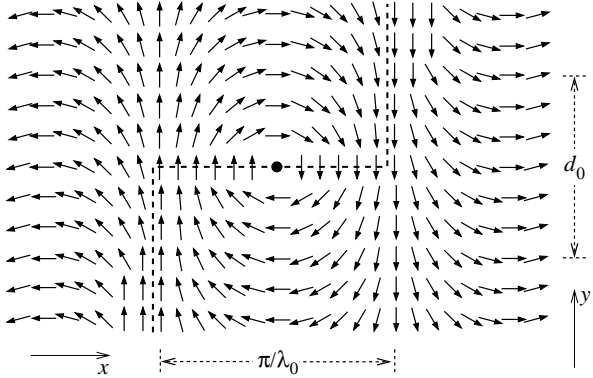,width=78mm,angle=-90}}
\vspace{2mm}
\caption{\label{fig:vdw} Elementary ``kink'' of the chiral domain wall
interface, corresponding to the jump of the wall by
$\pi/\lambda_{0}$. The arrows show the angle $\psi$; the vortex in $\psi$
corresponds to the two ``half-vortices'' in the fields
$\phi_{\pm}=\psi\mp\lambda_{0}x$ living at the opposite sides of the
boundary. The position of the domain wall is
indicated with the dashed line.}
\end{figure}

Further, it is easy to see that the chiral DW cannot move freely,
since the infinitesimal displacement of the DW coordinate $x_{DW}$
would cause global change of the phase $\psi$ at $x\to\infty$. The DW
can only ``jump'' by the integer multiples of $\pi/\lambda_{0}$, then the
phase at infinity changes by the integer multiples of $2\pi$. The jump by
$n\pi/\lambda_{0}$ involves formation of $n$ vortices bound on the DW,
the elementary $n=1$ jump is schematically shown in Fig.\
\ref{fig:vdw}. 
The energy per such a {\em bound vortex\/} can be estimated as
\begin{equation} 
\label{Ebv} 
E_{bv}\simeq\pi\sqrt{F}\ln(\pi/\lambda_{0})\,,\quad \lambda_{0}\ll 1\,.
\end{equation}

\begin{figure}
\mbox{\psfig{figure=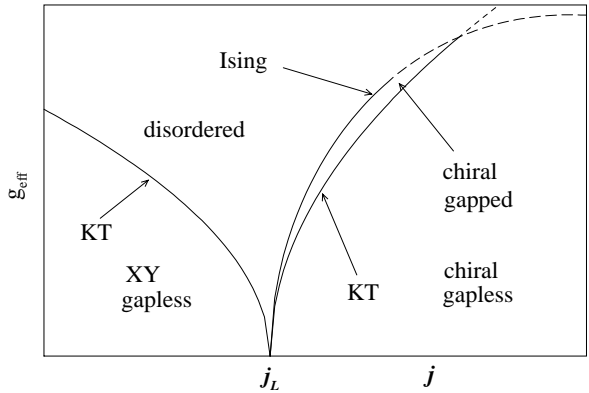,width=78mm,angle=-90}}
\vspace{2mm}
\caption{\label{fig:diag-heli} Sketch of the phase diagram of the 2D
helimagnet described by (\protect\ref{AE}).}
\end{figure}

Since the Ising transition is governed by the discrete fluctuations of
the DW interface, it is natural to use the so-called
M\"uller-Hartmann-Zittartz, or the ``solid-on-solid'' approximation
\cite{MHZ77}.  In this approach the transition temperature is
determined by looking for the point where the free energy $\sigma$
of the DW interface becomes zero; a simple calculation yields the
following equation for the critical coupling $g=g_{c}^{I}$:
\begin{equation} 
\label{gI} 
\sigma=E_{DW}-{g\over d_{0}}\ln\Big\{
1+d_{0}[\mbox{cotanh}{E_{bv}\over 2g}-1]\Big\}=0\,,
\end{equation}
where $d_{0}\simeq \pi/(\lambda_{0}\sqrt{F})$ is the characteristic
size of the bound vortex in the $y$ direction (in the derivation of
(\ref{gI}) we have assumed that the distance along the $y$ axis
between two successive  ``jumps'' should be greater than
$d_{0}$). 
This equation can be solved numerically, and 
for $j\leq 0.26$ the solution is
well fitted with the function $g_{c}^{I}\simeq 1.62
\lambda_{0}+0.28\lambda_{0}^{2}$, thus at $j\to j_{L}$ the Ising
transition temperature is larger than the KT one,
$g_{c}^{I}>g_{c}^{KT}\simeq {\pi\over2}\lambda_{0}$, confirming the
consistency of our assumption.

Away from the Lifshitz point the
above discussion of the Ising transition is no more valid, because the
characteristic size of the bound vortex and the DW thickness become
comparable with the lattice constant, and the continuum description
breaks down. However, it is known that $E_{DW}$ saturates at
$C_{DW}\approx0.87$ for $j\gtrsim 0.8$ \cite{Harada84}; one could also
speculate that for $j\to\infty$ the energy of the bound vortex
$E_{bv}\to C_{bv}\sqrt{j}$, where $C_{bv}$ is some constant, and then
from (\ref{gI}) one obtains $g_{c}^{I}\simeq C_{bv}\sqrt{j}/\ln j$ at
$j\to\infty$. On the other hand, according to (\ref{gKT})
$g_{c}^{KT}\to \pi\sqrt{j}$ in the same limit. Thus, one may expect
that above a certain critical value of $j$ the Ising transition
temperature $g_{c}^{I}$ becomes lower than $g_{c}^{KT}$, and the
gapped chiral phase disappears.

The resulting conjectured
phase diagram of the 2D helimagnet (\ref{AE}) is shown in Fig.\
\ref{fig:diag-heli}
It should be mentioned that our picture of the transitions
in 2D XY helimagnet disagrees strongly with that presented in
\cite{GarelDoniach80}. In the latter work, using the arguments of
\cite{Einhorn+80}, it was concluded that at low temperatures the
vortices are bound by strings, which would suppress the KT transition
and make the Ising transition to occur first with increasing the
temperature. However, the argument of \cite{Einhorn+80} is adequate
only for systems with broken {\em in-plane\/} symmetry, which is not
the case here. Another point is that in the description used in
\cite{GarelDoniach80} the fields $\phi^{\pm}$, measuring
the deviations from the two {\em different\/} possible helix states
with opposite chirality, are allowed to live and interact at the {\em
same\/} space-time point, which, to our opinion, is rather unphysical.

\begin{figure}
\mbox{\psfig{figure=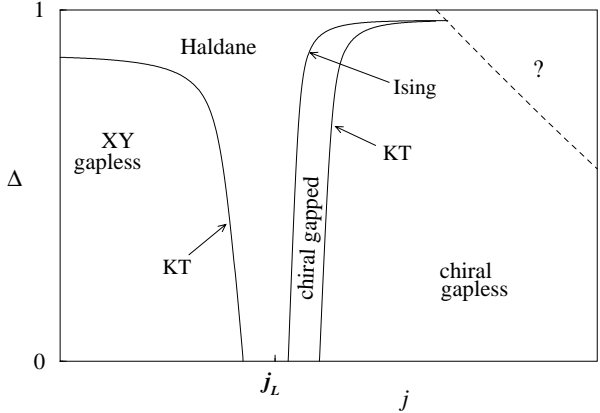,width=78mm,angle=-90}}
\vspace{2mm}
\caption{\label{fig:diag-chain} Schematic phase diagram of the anisotropic
frustrated spin chain (\protect\ref{ham}). 
}
\end{figure}

The above picture of the transitions in the 2D XY helimagnet is now
easily translated into the phase diagram of the frustrated spin chain,
which is schematically shown in Fig.\ \ref{fig:diag-chain} for $D=0$.
Very close to $j_{L}$, where $m_{0}\gg \Lambda$, which in terms of
$\varepsilon\equiv |j-j_{L}|$ and $\mu\equiv (1-\Delta)+{4\over3}D$
means $\varepsilon\ll 3\mu/(8\zeta\pi^{2})$, the renormalization of
the coupling constant is small, $g_{\rm eff}\simeq g_{0}$, and the
transition boundaries are approximately given by
\begin{equation} 
\label{crit1} 
\varepsilon_{c}^{a}={K_{a}\over\pi^{2}S^{2}}\big(D+{3+5\Delta\over4}\big)\,.
\end{equation}
Here the coefficient $K_{C:C}\simeq 1$ for the transition between
gapless and gapped chiral phases, $K_{C:H}\simeq 0.94$ for the
transition from the chiral gapped to the Haldane phase, and
$K_{H:XY}\simeq 2$ for the Haldane-XY transition. One can see that the
slope of the transition lines in the vicinity of $j_{L}$ is very large
(proportional to $S^{2}$), and for large $S$ the boundaries mover
closer and closer to the classical Lifshitz point $j=j_{L}={1\over4}$.

At larger deviations from $j=j_{L}$, when $m_{0}\ll \Lambda$, one has
the following equations for the phase boundaries:
\begin{equation} 
\label{crit2} 
\mu_{c}^{a}={8\zeta\pi^{2}\over3 }\varepsilon e^{-2\pi S\sqrt{\zeta} 
\big(\sqrt{\varepsilon}-\sqrt{\varepsilon_{0}^{a}}\big)}\,,\quad
\varepsilon_{0}^{a}={2K_{a}\over\pi^{2}S^{2}}\,,
\end{equation}
which are valid for $\sqrt{\varepsilon}-\sqrt\varepsilon_{0}^{a}\gg 1/S$.
One can see that the chiral gapped phase shrinks with increasing
$j$. It is interesting to note that for $j<j_{L}$ the Haldane phase is
stabilized by the frustration, in accordance with the numerical
results \cite{Tonegawa+95}.  Further away from $j_{L}$, when
$\lambda_{0}$ becomes of the order of $1$, the theory breaks down;
however, from the above arguments concerning the behavior of
$g_{c}^{I}$ we expect that the chiral gapped phase disappears above
certain critical value of $j$.

Certain limitations of the present theory should be mentioned.
Our approach does not distinguish between integer and half-integer $S$,
since we have integrated out the out-of-plane components, and the only
remaining topological charge, in-plane vorticity, plays no role.  The
topological term present in the full theory of the unit vector field
contains another quantum number, the so-called Pontryagin index; for
$j<j_{L}$ this term is known \cite{reviews2} to suppress the KT
transition for half-integer $S$, preventing the appearance of the
Haldane phase. At $j>j_{L}$ there is no topological term
\cite{RaoSen94,AllenSenechal95}, and one may expect that the KT
transition for $j>j_{L}$ survives also for half-integer $S$. However,
this point is not so clear since the ground state of a half-integer
spin chain at sufficiently strong frustration is spontaneously
dimerized \cite{Haldane82}, and our approach does not allow one to capture
this feature. Another limitation is that we cannot describe the hidden
(string) order in any way, and thus it is not possible to analyze the 
coexistence of the string order and chirality in the gapped chiral
phase observed in Refs.\ \cite{Kab+99,Hik+00} nor to study the transition to
the so-called double Haldane phase characterized by the absence of the
string order \cite{KRS96}.

{\em Acknowledgments.--\/} I would like to thank B.~A.~Ivanov and
H.-J.~Mikeska for fruitful discussions; the hospitality of Hannover
Institute for Theoretical Physics is gratefully acknowledged.  This
work was supported by the German Federal Ministry for Research and
Technology (BMBFT) under the contract 03MI5HAN5.

\end{multicols}
\end{document}